\documentstyle[twoside]{article}

\catcode`\@=11
\long\def\@makefntext#1{
\protect\noindent \hbox to 3.2pt {\hskip-.9pt  
$^{{\eightrm\@thefnmark}}$\hfil}#1\hfill}		

\def\@makefnmark{\hbox to 0pt{$^{\@thefnmark}$\hss}}	
	
\def\ps@myheadings{\let\@mkboth\@gobbletwo
\def\@oddhead{\hbox{}
\rightmark\hfil\eightrm\thepage}   
\def\@oddfoot{}\def\@evenhead{\eightrm\thepage\hfil
\leftmark\hbox{}}\def\@evenfoot{}
\def\sectionmark##1{}\def\subsectionmark##1{}}

\oddsidemargin=\evensidemargin
\newcounter{sectionc}\newcounter{subsectionc}\newcounter{subsubsectionc}
\renewcommand{\section}[1] {\vspace{12pt}\addtocounter{sectionc}{1} 
\setcounter{subsectionc}{0}\setcounter{subsubsectionc}{0}\noindent 
	{\tenbf\thesectionc. #1}\par\vspace{5pt}}
\renewcommand{\subsection}[1] {\vspace{12pt}\addtocounter{subsectionc}{1} 
	\setcounter{subsubsectionc}{0}\noindent 
	{\bf\thesectionc.\thesubsectionc. {\kern1pt \bfit #1}}\par\vspace{5pt}}
\renewcommand{\subsubsection}[1] {\vspace{12pt}\addtocounter{subsubsectionc}{1}
	\noindent{\tenrm\thesectionc.\thesubsectionc.\thesubsubsectionc.
	{\kern1pt \tenit #1}}\par\vspace{5pt}}

\newcounter{appendixc}
\newcounter{subappendixc}[appendixc]
\newcounter{subsubappendixc}[subappendixc]
\renewcommand{\thesubappendixc}{\Alph{appendixc}.\arabic{subappendixc}}
\renewcommand{\thesubsubappendixc}
	{\Alph{appendixc}.\arabic{subappendixc}.\arabic{subsubappendixc}}

\renewcommand{\appendix}[1] {\vspace{12pt}
        \refstepcounter{appendixc}
        \setcounter{figure}{0}
        \setcounter{table}{0}
        \setcounter{lemma}{0}
        \setcounter{theorem}{0}
        \setcounter{corollary}{0}
        \setcounter{definition}{0}
        \setcounter{equation}{0}
        \renewcommand{\thefigure}{\Alph{appendixc}.\arabic{figure}}
        \renewcommand{\thetable}{\Alph{appendixc}.\arabic{table}}
        \renewcommand{\theappendixc}{\Alph{appendixc}}
        \renewcommand{\thelemma}{\Alph{appendixc}.\arabic{lemma}}
        \renewcommand{\thetheorem}{\Alph{appendixc}.\arabic{theorem}}
        \renewcommand{\thedefinition}{\Alph{appendixc}.\arabic{definition}}
        \renewcommand{\thecorollary}{\Alph{appendixc}.\arabic{corollary}}
        \renewcommand{\theequation}{\Alph{appendixc}.\arabic{equation}}
        \noindent{\tenbf Appendix \theappendixc #1}\par\vspace{5pt}}
\newcommand{\subappendix}[1] {\vspace{12pt}
        \refstepcounter{subappendixc}
        \noindent{\bf Appendix \thesubappendixc. {\kern1pt \bfit #1}}
	\par\vspace{5pt}}
\newcommand{\subsubappendix}[1] {\vspace{12pt}
        \refstepcounter{subsubappendixc}
        \noindent{\rm Appendix \thesubsubappendixc. {\kern1pt \tenit #1}}
	\par\vspace{5pt}}

\newcommand{\textlineskip}{\baselineskip=13pt}
\newcommand{\smalllineskip}{\baselineskip=10pt}

\def\abstracts#1#2#3{{
	\centering{\begin{minipage}{4.5in}\baselineskip=10pt\footnotesize
	\parindent=0pt #1\par 
	\parindent=15pt #2\par
	\parindent=15pt #3
	\end{minipage}}\par}} 

\renewenvironment{thebibliography}[1]
	{\frenchspacing
	 \ninerm\baselineskip=11pt
	 \begin{list}{\arabic{enumi}.}
	{\usecounter{enumi}\setlength{\parsep}{0pt}
	 \setlength{\leftmargin 12.7pt}{\rightmargin 0pt} 
	 \setlength{\itemsep}{0pt} \settowidth
	{\labelwidth}{#1.}\sloppy}}{\end{list}}

\newcounter{itemlistc}
\newcounter{romanlistc}
\newcounter{alphlistc}
\newcounter{arabiclistc}

\newcommand{\fcaption}[1]{
        \refstepcounter{figure}
        \setbox\@tempboxa = \hbox{\footnotesize Fig.~\thefigure. #1}
        \ifdim \wd\@tempboxa > 5in
           {\begin{center}
        \parbox{5in}{\footnotesize\smalllineskip Fig.~\thefigure. #1}
            \end{center}}
        \else
             {\begin{center}
             {\footnotesize Fig.~\thefigure. #1}
              \end{center}}
        \fi}

\newcommand{\tcaption}[1]{
        \refstepcounter{table}
        \setbox\@tempboxa = \hbox{\footnotesize Table~\thetable. #1}
        \ifdim \wd\@tempboxa > 5in
           {\begin{center}
        \parbox{5in}{\footnotesize\smalllineskip Table~\thetable. #1}
            \end{center}}
        \else
             {\begin{center}
             {\footnotesize Table~\thetable. #1}
              \end{center}}
        \fi}

\def\@citex[#1]#2{\if@filesw\immediate\write\@auxout
	{\string\citation{#2}}\fi
\def\@citea{}\@cite{\@for\@citeb:=#2\do
	{\@citea\def\@citea{,}\@ifundefined
	{b@\@citeb}{{\bf ?}\@warning
	{Citation `\@citeb' on page \thepage \space undefined}}
	{\csname b@\@citeb\endcsname}}}{#1}}

\newif\if@cghi
\def\cite{\@cghitrue\@ifnextchar [{\@tempswatrue
	\@citex}{\@tempswafalse\@citex[]}}
\def\citelow{\@cghifalse\@ifnextchar [{\@tempswatrue
	\@citex}{\@tempswafalse\@citex[]}}
\def\@cite#1#2{{$\null^{#1}$\if@tempswa\typeout
	{IJCGA warning: optional citation argument 
	ignored: `#2'} \fi}}

\def\pmb#1{\setbox0=\hbox{#1}
	\kern-.025em\copy0\kern-\wd0
	\kern.05em\copy0\kern-\wd0
	\kern-.025em\raise.0433em\box0}

\def\fnt#1#2{\footnotetext{\kern-.3em
	{$^{\mbox{\scriptsize #1}}$}{#2}}}

\def\fpage#1{\begingroup
\voffset=.3in
\thispagestyle{empty}\begin{table}[b]\centerline{\footnotesize #1}
	\end{table}\endgroup}

\font\tenrm=cmr10
\font\tenit=cmti10 
\font\tenbf=cmbx10
\font\bfit=cmbxti10 at 10pt
\font\ninerm=cmr9

\font\eightrm=cmr8

\def\qed{\hbox{${\vcenter{\vbox{			
   \hrule height 0.4pt\hbox{\vrule width 0.4pt height 6pt
   \kern5pt\vrule width 0.4pt}\hrule height 0.4pt}}}$}}

\def\Dsl{\,\raise.15ex\hbox{/}\mkern-13.5mu D} 
\def\tr{{\rm tr}} 

\def\Gb{{G_{\bullet}}}

\def\Cb{{C_{\bullet}}}
\def\fb{f_{\bullet}} 
\def\Tb{T_{\bullet}}
 
\def\vol{{\rm vol}}

\def\p{\partial}

\def\CB{{\mathcal B}}

\def\CL {{\mathcal L}}
\def\CA {{\mathcal A}}
\def\CC{{\mathcal C}}
\def\CG{{\mathcal G}}
\def\CH {{\mathcal H}}

\def\IR{{\bf R} }
\def\IZ{{\bf Z} }
\def\sdtimes{\mathbin{\hbox{\hskip2pt\vrule height 4.1pt depth -.3pt width
.25pt
\hskip-2pt$\times$}}}
\def\p{\partial}
\def\half{{1\over 2}}

\begin{document}


\normalsize\textlineskip
\thispagestyle{empty}
\setcounter{page}{1}

\vspace*{0.88truein}

\fpage{1}
\centerline{\bf Anomalies, Gauss laws, and Page charges in M-theory }
\vspace*{0.37truein}
\centerline{\footnotesize Gregory W. Moore }
\vspace*{0.015truein}
\centerline{\footnotesize\it Department of Physics, 
Rutgers University,}
\baselineskip=10pt
\centerline{\footnotesize\it Piscataway, New Jersey, 08854-8019, USA}
\vspace*{10pt}
\vspace*{0.225truein}
 
\vspace*{0.21truein}
\abstracts{We review the $E_8$ model of the M-theory 3-form and its 
applications to anomaly cancellation, Gauss laws, quantization 
of Page charge, and the 5-brane partition function. We discuss 
the potentially problematic behavior of the model under parity.  
}{}{}


\vspace*{1pt}\textlineskip	
\section{Introduction}	
\vspace*{-0.5pt}
\noindent
 
In 1978  Cremmer,  Julia, and Scherk  found the action for 
11-dimensional supergravity \cite{Cremmer:1978km}. Twenty-six 
years later the theory has come to be regarded as a low energy limit of 
some hypothetical more fundamental ``M-theory.'' A satisfactory formulation 
of M-theory is still unknown. One set of clues to finding such a 
formulation lies in the  issues one encounters in 
formulating 11-dimensional supergravity in topologically nontrivial 
situations. While the   action principle in  \cite{Cremmer:1978km} is simple, it 
contains a very subtle  Chern-Simons term. In this note we review 
some recent work aimed at clarifying the mathematical nature of that 
term \cite{Witten:1996md,Witten:1996hc,Diaconescu:2000wz,Diaconescu:2000wy,Diaconescu:2003bm}. We will also describe 
briefly some related new results \cite{FM,pagepaper}. 
Another  motivation for this recent work is the clarification 
of anomaly cancellation issues in M-theory. This is discussed in 
section three below. A further motivation is the possibility that 
there are new topological terms 
in the action. (Such terms were found for type IIA supergravity 
in \cite{Diaconescu:2000wy} in exactly this way. For a general discussion 
see section 5.5 of \cite{FM}.)  As discussed in section 5, the Chern-Simons term also 
leads to a noncommutative structure in the theory leading to some 
important subtleties in flux quantization.  Finally, the 
considerations touched on here are of importance in understanding aspects of the 
M-theory 5-brane in topologically interesting situations. As we remark 
below, they are not without applications to currently fashionable topics.

While the $E_8$ formalism is very useful for studying some of the 
topological complexities of 11-dimensional supergravity it appears to 
have two important drawbacks.   The first is that the action of 
parity is subtle (and possibly impossible) to formulate in a satisfactory 
way. We describe some of the salient points in section 7 below.  
A second important challenge is the incorporation of 
elementary 5-branes in the formalism. In the spacetime external to the 
5-brane one finds a nontrivial $E_8$ instanton linking the 5-brane worldvolume.
However, to describe 5-branes we wish to include their worldvolume. Unfortunately, 
  the inclusion of nonzero magnetic current in the $E_8$ formalism 
presents an unsolved difficulty. 

\section{Defining the Chern-Simons term} 

Let $Y$ be an $11$-dimensional, oriented, spin manifold. In topologically 
trivial situations $M$-theory has an abelian gauge field, a globally 
defined 3-form $C \in \Omega^3(Y)$
with fieldstrength $G= dC \in \Omega^4(Y)$. 
\footnote{In general we follow the notation of \cite{Diaconescu:2003bm}.}
The exponentiated Euclidean action for the theory is (schematically): 
\begin{equation}
\exp\Biggl[ - 2\pi \int_Y {1\over \ell^9} \vol(g) R(g) + {1\over 2\ell^3} G\wedge *G  + \bar \psi \Dsl \psi \Biggr] 
\Phi(C)
\end{equation}
\begin{equation}
\Phi(C) =   
\exp\biggl(2\pi i \int_Y {1\over 6} C G^2 - C I_8(g) \biggr) 
\end{equation}
where $g$ is the metric, $\psi$ is the gravitino   and $\ell$ is the 11-dimensional Planck length. 
This form of the 
action cannot apply in topologically interesting situations in which 
the cohomology class $[G] \not=0$. 
If $\p Y = \emptyset$ the usual definition of a Chern-Simons 
term involves an extension to a bounding 12-manifold $Z$: 
\begin{equation}
\Phi(C) {\buildrel ? \over = } 
\exp\biggl(2\pi i \int_Z {1\over 6}  G^3 - G I_8(g) \biggr) 
\label{twelvedee}
\end{equation}
As it stands, this definition appears to depend on the extension. 
The existence of M2-branes implies 
$[G] = \bar a - \half \bar\lambda$ where $  \bar a\in \bar H^4(Y;\IZ)$ \cite{Witten:1996md}.
(The bar denotes reduction modulo torsion and $\lambda$ is the 
characteristic class of the spin-bundle on $Y$.) Thus the factor 
of $1/6$ looks problematic. In fact, since  
$[I_8(g)] = {p_2 - \lambda^2 \over 48}$, the definition 
(\ref{twelvedee}) appears to be ambiguous by a $96^{th}$ root of unity. 
It was pointed out by Witten in \cite{Witten:1996md} that $E_8$ 
index theory shows the situation is actually not that bad. 
Isomorphism classes of principal $E_8$ bundles on manifolds $M$ of dimension $\leq 15$ are  
in 1-1 correspondence with integral classes $a\in H^4(M,\IZ)$. 
Let  $P(a)$ denote an $E_8$ bundle with characteristic class $a\in H^4(Y,\IZ)$.   If
we identify $[G] = [\tr F^2 - \half \tr R^2] $, where $F$ is the fieldstrength 
of a connection $A$ on the bundle $P(a)$, then there is a remarkable identity 
\begin{equation}
{1\over 6} G^3 - G I_8 =  \biggl[{1\over 2} i(\Dsl_{A}) + {1\over 4} i(\Dsl_{RS}) \biggr]^{(12)}
\label{magic}
\end{equation} 
where $i(\Dsl)$ denotes the standard index density. The first term is 
for the Dirac operator coupled to $A$ in the adjoint representation 
while the second is for the Dirac operator coupled to $T^*Y - 4$. 
We extract the 12-form piece of the right hand side. 
(The simple formula (\ref{magic}) summarizes all the nontrivial 
group-theoretic identities used in Green-Schwarz anomaly cancellation 
for the $E_8 \times E_8$ theory, as well as the identities used by 
Horava and Witten \cite{Horava:1996ma}.)
Since the index is even in $12$-dimensions $\Phi(C)$ is in fact well-defined 
up to a sign. The sign ambiguity cannot be removed without introducing 
other fields.  See section 3 below.

\subsection{Boundaries} 

The extension to the case with boundary, $\p Y = X$ is nontrivial. It is best 
described by making  a choice of  a 
``model'' for the $C$-field. We will now explain what we mean by a ``model.'' 
The membrane coupling provides us with the 
{\it gauge equivalence class} of a $C$-field. Thus, an isomorphism class 
$[C]$ may be identified with a map 
\begin{equation}
[C]: \Sigma \rightarrow \exp(2\pi i \int_{\Sigma} C) 
\end{equation}
from the space of closed 3-cycles to $U(1)$, such that, if $\Sigma = \p B$ then 
$[C]: \Sigma \to \exp(2\pi i \int_B G)$. Mathematically, the membrane 
coupling is (a torsor for) the Cheeger-Simons group $\check H^4(Y)$.
\footnote{It is a torsor because of a shift in $C$ needed to cancel 
worldvolume anomalies on the membrane \cite{Witten:1996md}. This may also
be understood as being due to background magnetic current induced by $w_4$.}

While the mathematical formulation of the gauge equivalence class of a $C$-field 
is clear, there are different ways of expressing $C$ in terms of redundant 
variables. This issue does not arise in 
Yang-Mills theory, where there is a natural way: one uses the space of 
connections ${\rm Conn}(P)$ on a principal bundle $P$. In the case of the 
$C$-field, the language of categories turns out to be useful. This language 
applies to all gauge theories. 
 Abstractly the space of $C$-fields should be viewed as 
a groupoid, i.e. a category all of whose morphisms are invertible. 
 The gauge potentials are the objects, while the gauge transformations 
are the morphisms. The group of global gauge transformations is the automorphism 
group of the object. Different models for the $C$-field correspond to 
equivalent categories. In this note we use a particular model, the ``$E_8$ model 
for the $C$-field.'' Another model, based on the differential cohomology theory 
of Hopkins and Singer \cite{Hopkins:2002rd} is described in \cite{Diaconescu:2003bm} and is 
developed further in \cite{FreedHopkins}. 

In the $E_8$ model, a {\it ``$C$-field''  on $Y$  with characteristic class $a$ } 
is a pair $(A,c)$ in  
${\mathcal C}(Y) := {\rm Conn}(P(a)) \times \Omega^3(Y)$. The gauge invariant 
fieldstrength is $G = {\tr } F^2 - {1\over 2} {\tr} R^2 + dc $ 
so that, morally speaking, 
\begin{equation}
C   =CS(A) -{1\over 2} CS(g)+ c, 
\label{ceedef}
\end{equation}
can be written in terms of Chern-Simons forms. The objects of the groupoid 
are points in $\CC(Y)$. The morphisms are defined by a gauge group $\CG$, described in section 4 below. 
\footnote{In this definition we have fixed a bundle $P(a)$ for each $a$, at the cost of 
some naturality.  Section 3.5 of \cite{Diaconescu:2003bm} 
describes an equivalent category where no such choice is made.} 
  Note the metric dependence in $C$: 
  the space of 
bosonic fields in M-theory  is fibered over the space of metrics, the 
fiber is the space of $C$-fields. 

In the $E_8$ model we can write the Chern-Simons term of M-theory
as 
\begin{equation}
\Phi(C) =  \exp
\biggl[ 2\pi i \biggl\{{1\over 4}  \eta(\Dsl_A) + {1 \over 8} \eta(\Dsl_{RS})\biggr\} + 2\pi i I_{\rm local} \biggr] 
\label{elevendee}
\end{equation}
where $\eta$ is the Atiyah-Patodi-Singer invariant and 
\begin{equation}
I_{\rm local} =  
 \int_Y\left( c( {1\over 2} G^2-I_8)  -
{1\over 2}cdc  G +{1\over 6}c(dc)^2 \right).
\end{equation}

The importance of eq.(\ref{elevendee}) is that the formula is intrinsically formulated in 
11-dimensions and moreover the same formula holds on a manifold with boundary. 
However, we must now pay a price. $\Phi$ cannot be viewed as a $U(1)$ valued function 
but rather must be considered as a section of a line bundle 
${\mathcal L} \to  {\mathcal C}(Y) \times Met(Y)$. This line bundle has a connection. 
When $Y$ is closed the connection is only nontrivial in the metric directions, 
is flat,  and has $\pm 1$ holonomy on $Met(Y)/Diff^+(Y)$. When $\p Y = X$ 
is nonempty the connection has nontrivial components in the $C$-field 
directions. Heuristically $ {\mathcal A} =  2\pi \int_Y (\half G^2- I_8) \delta C $. 
An important point is that the curvature of $\CL$  is nonzero: 
\begin{equation}
{\mathcal F} = \pi \int_X G \delta C \delta C .
\end{equation}

\section{Anomaly cancellation and Setting the quantum integrand}

This section covers some new work done with D. Freed \cite{FM}.

Quite generally the quantum integrand of a path integral is a section of a 
line bundle $\CL_{qi}$ over the space of bosonic fields. This bundle 
is equipped with a connection $\nabla$.  In theories with fermions and/or Chern-Simons 
terms this line bundle with connection might well be nontrivial. If this is the 
case the path integral does not make sense - even formally - since one cannot 
add vectors in different lines. This is the geometrical interpretation of anomalies. 
In order to define a sensible path integral one must introduce a trivialization, 
i.e. a globally nonvanishing section ${\bf 1}$ of $\CL_{qi}$ so that, if $e^{-S}$
is the quantum integrand then $e^{-S}/{\bf 1}$ is a globally well-defined function, 
which can be integrated. 
Note that this requires that $\CL_{qi}$ be topologically trivial. Moreover, 
the connection $\nabla$ must be flat: This is the cancellation of local 
anomlies. Furthermore, the flat connection $\nabla$ must have no holonomy: 
This is the cancellation of global anomalies. In other words, in a well-defined 
theory $(\CL_{qi}, \nabla)$ must be {\it geometrically trivial}. Note that in 
an anomaly free theory there might still be a nontrivial choice of trivializing 
section ${\bf 1}$. In \cite{FM} this choice is called a ``setting of the 
quantum integrand.'' 

In the case of $M$-theory
both the $C$-field and the gravitino theories are quantum-mechanically 
inconsistent. That is, both $\Phi(C)$ and the gravitino partition function 
are sections of nontrivial line bundles with non-flat connections. However, it is 
shown in \cite{FM} that the tensor product is geometrically 
trivial. This is the Green-Schwarz anomaly cancellation. Moreover, it is shown 
in \cite{FM} that there is a canonical trivialization, thus leading to a 
canonical setting of the quantum integrand.

There are already anomalies in the case when $Y$ is closed.
This is the sign ambiguity mentioned below eq.(\ref{magic}). 
The gravitino partition function  
${\rm pf}(\Dsl_{RS})$ is a section of  the Pfaffian line bundle 
\begin{equation} L :={\rm PF}(\Dsl_{RS})\to Met(Y)
\end{equation}
   $L$ is a {\it complex line bundle} 
with  real structure. $L$ has a connection compatible with the 
real structure  so the  holonomies on $Met(Y)/Diff^+(Y)$ are $\pm 1$. In fact, 
the gravitino has a global anomaly. 
In \cite{FM}, sec. 2 one finds a 
natural geometric isomorphism $L \cong \CL$  leading to   global anomaly cancellation. That is, 
\begin{equation}
{\rm Pf}(\Dsl_{RS})\cdot \Phi
\end{equation}
is a well-defined function on $\CC(Y)\times Met(Y)/\CG\times Diff^+(Y)$.
This Green-Schwarz mechanism was already indicated in \cite{Witten:1996md}
and \cite{FM} establishes it rigorously.

When we consider anomaly cancellation on manifolds with boundary we  
need to distinguish temporal boundaries from spatial boundaries because 
of the boundary conditions which we will impose. In the case of temporal 
boundary conditions, we put global, or APS boundary conditions on the 
fermion fields. In this case, a  similar but rather more subtle story 
applies to establish the cancellation of Hamiltonian anomalies. 
 This is described in detail in \cite{FM}, sec. 4.2. 

We now consider spatial boundaries. 
With  local (i.e. chiral) boundary conditions   on fermions one can still define elliptic operators and 
study geometric invariants. ( D. Freed's student M. Scholl is studying a general class of 
local boundary conditions for Dirac operators on manifolds with boundary \cite{scholl}.)
Using these results one can produce a  geometric isomorphism between the line 
bundle of the Chern-Simons term and that of the fermion partition function. 
In this way one can give a rigorous  proof of anomaly cancellation in the Horava-Witten model. 
The advantage of this proof is that it covers simultaneously both local and global 
anomalies, and moreover  it becomes crystal clear that  
the anomaly   cancellation is completely {\it local}. (It has been pointed out that this issue is 
nontrivial \cite{Bilal:2003es}.) We are not being very precise here about the meaning of 
locality, but we note that the anomaly cancels boundary component by boundary component. 
In particular,  there is no topological obstruction to putting 
M-theory on an 11-manifold with any number of boundary components.  On each component 
we choose, arbitrarily, a sign   $\epsilon_i = \pm $ determining the chirality projection. 
Each component carries an independent $E_8$ super-Yang-Mills multiplet 
and we choose boundary conditions such that  $G\vert_{X_i}  = \epsilon_i ( \tr F^2(A_i) - \half \tr R^2(g_i)) $. 
There are a number of subtle details one encounters in checking this cancellation. 
Perhaps the most surprising is that, in some circumstances, the {\it Pfaffian} line bundle 
admits, globally, a well-defined square root. Again, for the many details we refer to \cite{FM}, sec. 4.3. 

The existence of these topological sectors of M-theory raises the interesting question of whether there 
are {\it solutions} of the equations of motion on manifolds of this type. This curious question 
remains open. 

\section{The Gauss law} 

Our next goal is to write the Gauss law for $C$-field gauge invariance. 
In the $E_8$ model $C= (A,c)$. Small gauge transformations act by $c \to c+ d \Lambda$, $\Lambda \in \Omega^2(Y)$.
It is usually said that gauge transformations are $c \to c+ \omega$, \quad $\omega \in \Omega^3_{\IZ}(Y)$. 
\footnote{$\Omega^p_{\IZ}(M)$ denotes the space of $p$ forms on $M$ with integral periods. Such forms are 
necessarily closed. } 
However, this does not properly account for global gauge transformations ``$\Lambda\sim $ constant.'' 
The correct choice is to replace $\Omega^3_{\IZ}(Y)$ by the Cheeger-Simons group $\check H^3(Y)$, and 
interpret $\omega$ as the fieldstrength of the differential character. 
What we stress here is that the Cheeger-Simons group $\check H^3(Y)$  is an extension: 
\begin{equation}
0 \to H^2(Y,U(1)) \to \check H^3(Y)  \to \Omega^3_{\IZ}(Y) \to 0 .
\label{realgg}
\end{equation}
We interpret $H^2(Y,U(1))$ as the group of  global gauge transformations. In the categorical 
language, these are the automorphisms of the object:
If $\alpha\in H^2(Y,U(1))$ then 
$\gamma_{\alpha}: (A,c) \to (A,c) $. But the automorphism   still has nontrivial physical effects. 
Firstly, it has a nontrivial effect on {\it open} membrane amplitudes. A second nontrivial 
effect emerges in   the formulation of the Gauss law for the gauge group $\CG$. 
\footnote{Actually, $\CG= \Omega^1({\rm ad} P) \sdtimes \check H^3(Y)$, where the first factor shifts $A \to A+ \alpha$, 
see \cite{Diaconescu:2003bm}.}

The Gauss law is the statement that physical wavefunctions of the $C$-field  must be gauge invariant: 
\begin{equation}
\gamma \cdot \Psi(  C) = \Psi( \gamma \cdot  C) \qquad \forall \gamma\in \CG, C \in \CC(X)
\end{equation}
Now the wavefunction is a section of the line $\CL$ in which $\Phi$ is valued. Thus, to  
 formulate the Gauss law we must define a lift: 
\begin{equation}
\matrix{
\CL& \qquad & {\buildrel \CG\over \rightarrow} & \qquad \CL \cr
    &        &                                  &     \cr
\downarrow &  &  &  \downarrow \cr
     &  &  & \cr
 \CC(X) & \qquad & {\buildrel \CG\over \rightarrow} & \qquad  \CC(X) \cr}
\end{equation}
To define the lift we combine the parallel transport using the connection on 
$\CL$ with a cocycle for the group action: 
\begin{equation}
\gamma\cdot \Psi(C) = \varphi(C, \gamma)^*\cdot \exp( \int_{C}^{\gamma\cdot C}  \CA )\cdot \Psi 
\end{equation}
where $\varphi(C, \gamma)$ is a cocycle, that is 
\begin{equation}
\varphi(C, \gamma_1) \varphi(\gamma_1\cdot C, \gamma_2) = e^{-i \pi \int_X G \omega_1 \omega_2} 
\varphi(C, \gamma_1 \gamma_2) 
\label{cocycle}
\end{equation}
where $\gamma_1,\gamma_2\in \CG$ are $C$-field gauge transformations with fieldstrength $\omega_1,\omega_2\in \Omega^3_{\IZ}(X)$. 
We will refer to $\varphi(C, \gamma)$  as the ``lifting phase.''  
Following a construction in \cite{Witten:1996hc} (described more fully in \cite{Diaconescu:2003bm}), 
given  $C\in \CC(X)$ and  $ \gamma \in \CG $ we construct a twisted $C$-field $C_{\gamma}$ on $Y= X\times S^1$: 
$C_{\gamma}(x,1) = \gamma\cdot C_{\gamma}(x,0) = \gamma \cdot C
$. Then we define 
$
\varphi(C,\gamma):= \Phi(C_{\gamma}) 
$. 

Since (\ref{realgg}) is an extension the Gauss law consists of two 
statements. For $\gamma=\gamma_\alpha$, $\alpha\in H^2(X,U(1))$ we obtain the electric 
charge tadpole condition. Once this law is satisfied we can study the 
Gauss law for $\gamma\in \Omega^3_{\IZ}(X)$. This leads to the quantization of 
``Page charge.'' 

The tadpole condition has been described in detail in \cite{Diaconescu:2003bm}. 
Assume $X$ is compact. 
A global gauge transformation  $\gamma_\alpha$, $\alpha\in H^2(X,\IR/\IZ)$ 
acts nontrivially on quantum wavefunctions. 
If $\Psi \in \CL_{A,c}$ then 
$
\gamma_\alpha \cdot \Psi =\exp\bigl[2\pi i \langle Q ,\alpha \rangle \bigr] 
 ~ \Psi 
$, where $Q \in H^8(X,\IZ)$ is the $C$-field electric charge. 
Thus, if $Q\not=0$ then $\Psi=0$. 
From the definition of the group lift we get a formula for $Q$. 
It only depends on the characteristic class $a$, so we may write $Q(a)$. 
It is easy to show that   
$
\bar Q = [\half G^2 - I_8]_{DR}
$
thus recovering the usual condition of \cite{Sethi:1996es}. Nevertheless, 
$Q(a)$ is an {\it integral refinement} of $[\half G^2 - I_8]= \half \bar a (\bar a - \bar \lambda) + 30 \hat A_8$, 
and hence $Q(a)=0$ carries further information related to torsion.  
Not much is known about $Q(a)$. It is a 
quadratic refinement of the cup product. This and some other 
pertinant facts can be found in \cite{Diaconescu:2003bm}. 

When $Q =0$ we can have nonzero gauge invariant wavefunctions 
$\Psi(C)\in \Gamma(\CL) $. There is still further information in 
the statement of gauge invariance. In order to demonstrate the 
physical interpretation it is convenient to trivialize $\CL$. 
This entails choosing a basepoint so $C = \Cb+c$, and replacing 
the wavesection $\Psi(C)$ by a wavefunction $\psi(c)$. The result of 
a careful analysis \cite{pagepaper} is that the Gauss law may be written: 
\begin{equation}
\psi(c+\omega) = e_\omega(c) \psi(c) \qquad\qquad \forall \omega \in \Omega^3_{\IZ}(X)
\label{glaw}
\end{equation}
where 
\begin{equation}
e_{\omega}(c) := \varphi(\Cb, \omega)^* e^{2\pi i \int_X \left(
\half \Gb +  {1\over 6} d c\right)  c \omega }.
\label{efactor}
\end{equation}

\section{Page Charges}

Equation eq.(\ref{glaw}) can be interpreted physically  by rewriting it in the form 
\begin{equation}
\exp\bigl( 2\pi i \int_X \omega P \bigr) \psi =  \fb(\omega) \psi \qquad \forall \omega \in \Omega^3_{\IZ}(X)
\label{plaw}
\end{equation}
where $P$ is an operator-valued $7$-form. In order to prove this one 
notes that on spin 10-manifolds the cocycle in (\ref{cocycle}) is in 
fact $\IZ_2$-valued (This is nontrivial since $[G]$ has half-integer periods). 
Then it follows that $\varphi(\Cb, \omega)$ is 
{\it linear} on $\Omega^3_{2\IZ}$ and hence of the form 
$\varphi(\Cb,\omega) = \exp[2\pi i \int \omega \Tb]$. The $7$-form  $\Tb\in \Omega^7(X)$
is a trivialization $dT_\bullet = \half \Gb^2 - I_8$. It is only defined modulo a 
form with half-integer periods. We make a definite choice and define 
$\fb(\omega):= \varphi(\Cb,\omega)^* e^{2\pi i \int \omega \Tb}$ for all $\omega\in \Omega^3_{\IZ}(X)$. 
 This is a 
$\IZ_2$-valued cocycle satisfying (\ref{cocycle}). It is then elementary to show 
that  (\ref{glaw}) is equivalent to (\ref{plaw}) provided
\begin{equation}
 P = {1\over 2\pi }  \Pi + \bigl( \half \Gb c + {1\over 6}  cdc \bigr) + T_\bullet.
\label{pagecharge}
\end{equation}
where $\Pi$ is the canonical momentum of $c$. 
%
The expression  (\ref{pagecharge}) is nothing other than 
the ``Page charge'' of supergravity, formulated in 
the canonical formalism. This 7-form flux should be considered as the 
electro-magnetic dual of the flux $G$. Morally speaking, $P = dC_6$ where $C_6$ is 
the $6$-form potential that couples to the 5-brane. 

We are now in a position to study the quantization of Page charge. Here we encounter a 
surprise. If $[G]=0$,  the 
quantum Gauss law for large $C$-field gauge transformations implies 
$
[P] \in \bar H^7(X;\IZ)
$.
This is the naive electro-magnetic dual to the naive
quantization of magnetic flux: 
$[G] \in \bar H^4(X;\IZ)$.
However, when $[G]\not=0$, things are quite different. 
For $\phi \in H^3_{DR}(X)$ define 
$
P(\phi) := \int_X \phi \wedge P 
$.
An easy computation shows that 
\begin{equation}   
[P(\phi_1), P(\phi_2) ] = {i \over 2\pi } \int \phi_1 \wedge \phi_2 \wedge G .
\label{noncom}
\end{equation} 
Equation eq.(\ref{noncom}) is important. It means, first of all, that not all $P(\phi)$ can be 
simultaneously diagonalized. Moreover, $[P]$ is not even gauge invariant. If $U(\omega) := \exp[2\pi i \int \omega P ] $
implements large gauge transformations then (as was noted in a special case in \cite{Beasley:2002db}) 
\begin{equation}
U(\omega) P(\phi) U(\omega)^{-1} = P(\phi)  - \int \omega \phi G. 
\end{equation}

In general, the conserved gauge invariant ``Page charges'' or electric fluxes 
 should be regarded as characters of a certain group
which we will call the {\it magnetic translation group. }
When $[G]=0$ this group is simply $H^3(X,U(1))$, and hence we recover the lattice 
of fluxes, $H^7(X,\IZ)$. In general,  with $[G]\not=0$, the group is generated by the 
gauge invariant  operators  
$
W(\phi) := e^{2\pi i P(\phi)} 
$
where   $\phi$ is such that: 
$\int \phi \omega G \in \IZ $ for all $  \omega \in H^3(X,\IZ)$.
 Note that the group is 
in general nonabelian: 
\begin{equation}
 W(\phi_1) W(\phi_2) = e^{-i \pi \int \phi_1 \phi_2 G} W(\phi_1 + \phi_2) = e^{-2\pi i \int \phi_1 \phi_2 G} W(\phi_2) W(\phi_1) .
\end{equation}
In summary, the naive lattice of (magnetic,electric) fluxes $H^4(X,\IZ)\oplus H^7(X,\IZ)$ is modified in 
two ways. The first factor is constrained by the tadpole constraint $Q(a)=0$. The second factor is 
replaced by the character group of the magnetic translation group.

A comparison with ordinary gauge theory might help in understanding better what is going on here. 
Consider $U(1)$ gauge theory on spacetimes of the form $X \times \IR$, where $X$ is an $n$-dimensional 
Riemannian manifold. If we take the action $S = \int_{X \times \IR} -{1\over 2 e^2} F*F$ then 
the Hilbert space of the theory is graded by $H^2(X,\IZ) \oplus H^{n-1}(X,\IZ)$. 
The first component is $c_1$ of the line bundle on which $A$ is a connection, 
while the second component is the quantized electric flux. This grading can be understood 
elegantly as follows. 
\footnote{Thanks to G. Segal for some illuminating remarks.}  The space of gauge 
equivalence classes of line bundles with connection on $X$ is 
the Cheeger-Simons group $\check H^2(X)$, and therefore the Hilbert space is - formally - $L^2(\check H^2(X))$. 
Now, note that $\check H^2(X)$ is an abelian group. Quite generally, if $A$ is an abelian group then a Heisenberg extension of 
$A \times \hat A$ acts on $L^2(A)$ where $\hat A$ is the group of characters of $A$. If  $X$ is oriented the Poincar\'e dual group
to $\check H^2(X)$  is $\check H^{n-1}(X)$. The subgroup $H^1(X,U(1)) \times H^{n-2}(X,U(1))$ of $A \times \hat A$ 
acts on Hilbert space with trivial extension. The characters of this subgroup are simply 
$H^2(X,\IZ) \oplus H^{n-1}(X,\IZ)$. Now, let us   consider 3d massive abelian gauge theory with action 
\begin{equation}
S = \int_{\Sigma \times \IR} - {1\over 2e^2} F*F + 2\pi  \int_{\Sigma \times \IR} k A dA 
\label{threedee}
\end{equation}
where $\Sigma$ is a Riemann surface. The exponentiated Chern-Simons term must be considered as 
a section of a line bundle $\CL_k \to \check H^2(\Sigma)$. We now identify the 
Hilbert space as a space of $L^2$ sections $\Gamma(\check H^2(\Sigma); \CL_k)$. 
The wavefunction is only nonzero on the component with $c_1=0$ (this is the analog of the tadpole condition $Q(a)=0$ above). 
Moreover, because $\CL_k$ carries a nontrivial connection the translation symmetry is broken 
and replaced by a   Heisenberg group extension of 
$H^1(\Sigma, \IZ/k\IZ)$. In the analogy with Chern-Simons theory $k$ corresponds to $\half [G]$ 
and the (noncommuting) Wilson line operators correspond to the operators $W(\phi)$. 

We expect that the above remarks will have some important implications for the classification of 
RR fluxes in type II string theory. It is commonly believed that the topological 
sectors are classified by twisted K-theory. (See \cite{Witten:2000cn,Moore:2003vf} for  
recent reviews.) Naively one might expect the classification of RR fluxes in the 
background of a nontorsion $H$ field to be given in terms of the image of the 
Chern-character of twisted K-theory \cite{Mathai:2002yk}, analogous to the quantization condition 
proposed in \cite{Moore:1999gb,Freed:2000tt}. A discussion of this proposal (and other relevant 
matters) can be found in \cite{Mathai:2003mu}. Dimensional reduction of the above formulae indicate that 
the situation is more complex and needs further investigation.

The phenomenon we have described is probably closely related to the Hanany-Witten effect
\cite{Hanany:1996ie} and to the noncommuting brane charges of \cite{Gukov:1998kn}.
Similar noncommutative structures have appeared in compactifications of M-theory on 
tori \cite{Obers:1997kk} and in formulations of  M-theory using  
the $C$-field together with its electromagnetic dual   \cite{Cremmer:1998px}.

\section{Application: The 5-brane partition function}

In the 3D Chern-Simons theory of eq.(\ref{threedee})  the 
 dynamics of the topological (flat) modes of $A$ is that of an electron 
on a torus $H^1(\Sigma; U(1))$ in a constant magnetic field. In a long 
distance approximation of M-theory, ``$\ell \to 0$,'' where $\ell$ is the 11-dimensional Planck length  
one only keeps the harmonic modes of the $C$-field and an analogous story holds.  If we introduce a basis $\omega^a$ of 
the space $\CH^3(X)$ of harmonic 3-forms on $X$ then we may expand    $c= \sum_a c_a \omega^a$, 
and the effective Hamiltonian for these modes may be shown to be
\begin{equation} 
H_{\rm eff} = h_{ab} \bigl( - i {\p \over \p c_a } - \pi \CB^{aa'} c_{a'} \bigr) \bigl( - i {\p \over \p c_b } - \pi \CB^{bb'} c_{b'} \bigr) 
\label{hamiltonian}
\end{equation}
where $h^{ab} = \int_X \omega^a * \omega^b $ and the ``magnetic field'' is $\CB^{ab} 
= \int_X G \omega^a \omega^b$. We effectively have a Landau-level problem on the torus $H^3(X,\IR)/\bar H^3(X,\IZ)$. 
The Page charge operator corresponds to the magnetic translation operator. 

As an application, we can use the above formalism to derive Witten's prescription for the 
5-brane partition function \cite{Witten:1996hc}.  In the process of doing so we will underscore a point which 
is almost always misunderstood in the literature. 
Our approach will be via the AdS/CFT correspondence. We consider 
$X= D \times S^4$, where $D$ is a compact $6$-fold, so $X$ is a  
 conformal boundary at infinity for an asymptotically AdS space $Y$: 
\begin{equation}
ds^2 \to (k^{2/3} \ell^2) \biggl[ dr^2 + e^{2r} ds^2_{D} + {1\over 4} ds^2_{S^4} \biggr] ,
\end{equation} 
and   $G \to G_{\infty} = k \omega_{S^4} + \tilde G $, where $\tilde G \in \Omega^4(D)$. 
According to AdS/CFT for $k\gg 1$ the partition function of M-theory on $Y$ is the 
partition function of the $U(k)$ $(2,0)$ theory on $D$. Now $U(k) = {SU(k) \times U(1) \over \IZ_k} $ 
where the $U(1)$ couples to the center of mass degree of freedom of the 5-branes. This couples 
to the harmonic modes  of $c$ at infinity (for simplicity we   denote these as $c$) and, contrary to what is usually stated, 
does not completely decouple. In fact, the partition function of the $(2,0)$ theory may be written 
as
\begin{equation}
Z\bigl[U(k)\qquad (2,0)-{\rm theory}\bigr]
= \sum_{\beta \in \Lambda_1/k \Lambda_1} \zeta^\beta \Psi_\beta(c ) 
\label{singleton}
\end{equation}
where $H^3(D,\IZ) = \Lambda_1 \oplus \Lambda_2$ is a Lagrangian decomposition of $H^3(D,\IZ)$ with its canonical symplectic structure. 
(For a discussion of similar decompositions in $AdS_3$ and $AdS_5$ see \cite{Witten:1998wy,Maldacena:2001ss,Gukov:2004id}.) 
In eq.(\ref{singleton}) 
$\zeta^\beta$ is the contribution of the $SU(k)/\IZ_k$ $(0,2)$ theory. As pointed out in \cite{Witten:1998wy}, 
$\beta$ should be considered as a label for the 't Hooft sectors of the $SU(k)/\IZ_k$ $(0,2)$ theory. 
(Note that for $D = D' \times S^1$, the theory reduces to   $SU(k)/\IZ_k$ gauge theory on $D'$ and we have 
a natural symplectic splitting with 
$
\Lambda_1 = H^2(D', \IZ) $, but this is precisely the group classifying 't Hooft sectors.)   
On the other hand, the magnetic translation group is a Heisenberg group extending 
$H^3(D,\IZ_k)$ and  the formula for $\Psi_\beta$ given below makes it clear that 
\begin{equation}
W(\phi_1) \Psi_\beta  = \Psi_{\beta + \phi_1} \quad \qquad \qquad \qquad \phi_1\in \Lambda_1/k\Lambda_1 
\end{equation}
\begin{equation}
W(\phi_2) \Psi_\beta  = e^{2\pi i k \langle \phi_2, \beta\rangle} \Psi_\beta \ \qquad \qquad \phi_2\in \Lambda_2/k\Lambda_2
\end{equation}
giving the standard representation of the Heisenberg group. Thus, the 't Hooft sector label is AdS/CFT dual to the 
Page charge. 

Let us now come to the explicit formulae for the conformal blocks of the 5-brane theory. 
To derive the 5-brane partition function, in the $\ell \to 0$ approximation, we solve 
for the eigenstates of eq.(\ref{hamiltonian}). The ground state on $\CH^3(X)$ is the lowest Landau level.
We may take $c\in \CH^3(D)$, and then an  overcomplete basis of wavefunctions has the form 
$\Psi_v(c) = e^{-{\pi k \over 2} \int_{D} c*c + \int_D v   (1+i*)c }$. Here $v \in \CH^3(D)$, 
and the Landau level is infinitely degenerate. 
However, we must project these 
wavefunctions onto gauge invariant states, so we average over large gauge transformations: 
\begin{equation}
\overline{\Psi}_v =  \sum_{\omega\in \CH^3_{\IZ}(D)} (e_\omega(c))^* \Psi_v(c+\omega) 
\end{equation}
where $e_\omega(c)$ was defined in eq.(\ref{efactor}). 
Written out explicitly this becomes 
\begin{equation}
\overline{\Psi}_v = \sum_{\omega\in \CH^3_{\IZ}(D) }  \varphi(\check \Cb, \omega) 
 \exp\biggl\{ - {\pi k \over 2}  \int_{D} (c+\omega)*(c+\omega)
 - i \pi k \int_{D} c \wedge \omega \biggr\}\exp\biggl\{ \int_{D} v \wedge (1+i*)(c+\omega) \biggr\}
\label{5branesum}
\end{equation}
The span of these wavefunctions is finite-dimensional, as is most easily seen by 
 performing a Poisson resummation with respect to $\Lambda_2$. One then obtains 
\begin{equation} 
\overline{\Psi}_v = \sum_{\beta \in \Lambda_1/k\Lambda_1} {\Psi}_{\beta}(c)  {\Psi}_\beta(v)^*
\label{holofactor}
\end{equation}
where $\Psi_\beta(c) = e^{Q} \Theta_{\beta,k/2}$ with $Q$ a quadratic (nonholomorphic) form in $c$ and 
$\Theta_{\beta,k/2}$ a holomorphic level $k/2$ theta function. Holomorphy refers to the complex structure 
on $\CH^3(D)$ defined by Hodge $*$ \cite{Witten:1996hc}. The argument of the theta function is 
shifted by characteristics, which can be deduced from $\varphi(\Cb, \omega)$. 
 In this way one derives explicit formulae for the conformal blocks.

We recognize in the sum and the first exponential in eq.(\ref{5branesum}) the 5-brane partition function 
of Witten \cite{Witten:1996hc}. The sum over $\omega$ is therefore interpreted as a sum over 
instantons for the chiral 2-form on the 5-brane. Of course, our derivation is only valid 
for $k\gg 1$, but we expect that the formulae hold for all values of $k$. In particular, for $k=1$ 
(\ref{holofactor}) is a holomorphic square. 
Note the inclusion of the lifting phase 
$\varphi(\check \Cb, \omega)$. Without this phase, Poisson resummation will not produce theta 
functions of the correct level, or with the correct characteristics. In particular, without 
the phase one finds a sum over level $2k$ theta functions.  Moreover, the lifting 
phase shows that the characteristics of the theta function depend on the metric. 
Indeed, one can show that if we change the metric, holding $(A,c)$ fixed then 
\begin{equation}
{\varphi(C_{\bullet,1}, \omega) \over \varphi(C_{\bullet,2}, \omega)} = \exp[2\pi i k \int_{D} \omega CS( g_1,  g_2) ] 
\end{equation} 
where $CS(g_1,g_2)$ is the relative Chern-Simons form for the two metrics. 
There are also potential contributions to the characteristics from quantum 
corrections to the Born-Oppenheimer approximation one uses when separating 
harmonic from nonharmonic modes of the $c$-field.

The issue of characteristics can be important in applications, such as 5-brane 
instantons. A theta function with characteristics has an expansion schematically 
of the form $\Theta \sim q^{\theta^2/2} + \cdots $. Thus, if $q$ is small (e.g. 
because some coupling is weak) and $\theta$ is nonzero, there can be suppression of 
5-brane instanton amplitudes. Such suppressions can have consequences. For example, 
using these considerations it might be possible to derive an interesting  
 lower bound on the values of the string coupling for 
which the constructions of \cite{Denef:2004dm} are self-consistent.

\section{The problem with parity}

M-theory is parity invariant, and should in principle be formulated in 
a way which makes sense on unoriented, and possibly nonorientable, manifolds. 
The formalism described above makes heavy use of an orientation on $Y$. 
Extending the $E_8$ formalism to a parity invariant formalism is subtle and 
potentially problematic. 
\footnote{This section is based on discussions with D. Freed.}
 There is no difficulty at all describing the 
action of parity on isomorphism classes of the $C$-field. We take $[C]^P = -[C]$, that is, 
any parity transform $C\to C^P$ must satisfy
\begin{equation}
\exp  2\pi i \int_{\Sigma} C^P   = \bigl( \exp  2\pi i \int_{\Sigma} C \bigr)^* .
\label{paritydef} 
\end{equation} 
Note that $G^P = - G$ and $a^P = \lambda-a$. 
In the $E_8$ model we understand $C$ in this equation as in eq.(\ref{ceedef}). 
However, there is 
no natural way to map $A\in {\rm Conn}(P(a))$ to $A^P\in {\rm Conn}(P(\lambda-a))$. 
By contrast, in the  rival model \cite{Diaconescu:2003bm,FreedHopkins} 
based on Hopkins-Singer cocycles the action of parity is simple and natural. 
 In the latter model a $C$-field is 
represented by a triple $(a,h,G) \in \check{C}(Y):=C^4(Y,\IZ)\times C^3(Y,\IR)\times \Omega^4(Y)$ 
and parity is simply the transformation $(a,h,G) \to (\lambda(g) - a, -h, -G)$
(there is a functorial  choice of a representative $\lambda(g)$ of the class $\lambda\in H^4(Y,\IZ)$).
This presents a serious problem for the $E_8$ model. It can be traced to the fact that there is a 
natural group structure on $\check{C}(Y)$, but there is no natural group structure on 
$\amalg_a  {\rm Conn}(P(a))$.

One way to address the parity problem was discussed in \cite{Diaconescu:2003bm}. Let $Y_d$ be 
the orientation double cover of $Y$ and let $\sigma$ be the Deck transformation 
so that $Y_d/\langle \sigma \rangle = Y$. We then define a ``parity invariant 
$C$-field on $Y$'' to be a $C$-field on $Y_d$ such that $\sigma^*[C] = [C]^P$. 
If $Y$ is orientable this definition amounts to defining a parity invariant $C$-field 
on $Y$ as a pair of ordinary $C$-fields on $Y$, namely, $\bigl( (A,c), (A',c')\bigr)$ such that 
\begin{equation} 
\exp 2\pi i \int_{\Sigma} C = \bigl(\exp 2\pi i \int_{\Sigma} C'\bigr)^* 
\label{parity} 
\end{equation}
The morphisms of the groupoid are simply $\CG \times \CG$. The space of isomorphism 
classes is the same as before. However, at  this point we 
encounter a new problem: The automorphism group of an object in our 
new groupoid is $H^2(Y,U(1)) \times H^2(Y, U(1))$ and hence the 
groupoid is {\it inequivalent} to the previous one, even when $Y$ is orientable!
 A potential solution to this  difficulty is that one must require (\ref{parity}) hold for open 
membrane worldvolumes $\Sigma$. Such a constraint reduces the automorphism 
group to a single copy of $H^2(Y,U(1))$, as desired, but introduces yet another 
difficulty. For open membranes, the left and right hand sides of (\ref{parity}) 
are sections of line bundles (over the space of $2$-cycles in $Y$). These line 
bundles are isomorphic, but not naturally so. The set of isomorphisms is a 
torsor for $H^2(Y,U(1))$, which accounts for the ``second'' copy in the automorphism 
group of an object in our parity-invariant groupoid.   Fortunately, this extra factor of $H^2(Y,U(1))$ appears to 
have no physical effect, and hence we effectively have an equivalent groupoid. 
Thus, in the author's current opinion, the parity invariant $C$-field model is physically 
viable. However, this issue clearly deserves further scrutiny.

Note that the above formulation of the $E_8$ model has the elegant consequence that 
the underlying topological gauge group is $E_8 \times E_8$ when $Y$ is 
orientable, while it is simply a single copy of $E_8$ when $Y$ is nonorientable.



\section{Acknowledgements}

The author would like to thank E. Diaconescu and D. Freed for collaboration on these 
matters, and D. Freed for useful comments on the draft. 
 He is also indebted to N. Lambert,  G. Segal and E. Witten for important discussions. 
The author would like to thank the theory group at the LPTHE, Jussieu and the 
Aspen Center for Physics for hospitality during the course of part of this work. 
This work was supported  in part by DOE grant DE-FG02-96ER40949.

\end{document}